\documentclass{osa-article}
\journal{osac}

\articletype{Research Article}

\begin{document}

\title{Hyperparallel transistor, router and dynamic random access memory  with unity fidelities}

\author{Ji-Zhen Liu\authormark{1}, Ning-Yang Chen\authormark{1}, Wen-Qiang Liu\authormark{1}, Hai-Rui Wei\authormark{1*}and Ming Hua\authormark{2}}

\address{\authormark{1}School of Mathematics and Physics, University of Science and Technology Beijing, Beijing 100083, China\\
\authormark{2} Department of Applied Physics, School of Science, Tianjin Polytechnic University, Tianjin 300387, China }

\email{\authormark{*} hrwei@ustb.edu.cn} 



\begin{abstract}
We theoretically implement some hyperparallel optical elements, including quantum single photon transistor, router, and dynamic random access memory (DRAM). The inevitable side leakage and the imperfect birefringence of the quantum dot (QD)-cavity mediates are taken into account, and unity fidelities of our optical elements can be achieved. The hyperparallel constructions are based on polarization and spatial degrees of freedom (DOFs) of the photon to increase the parallel efficiency, improve the capacity of channel, save the quantum resources, reduce the operation time, and decrease the environment noises. Moreover, the practical schemes are robust against the side leakage and the coupling strength limitation in the microcavities.
\end{abstract}


\section{Introduction}\label{sec1}

By exploiting the superposition principle, quantum information processing (QIP)  offers great advantages over the classical information processing in factoring power, security, discrete logarithms, efficient simulation, and modeling \cite{book}. Different from the normal QIP that is only acting on single degree of freedom (DOF) \cite{Ai2,Ai3}, the hyperparallel QIP is performing on more than one independent DOFs, simultaneously. Hyperparallel QIP has been shown decisive advantages in improving its encoding capacity, lowing loss rate, reducing experimental requirements, and decreasing affection by decoherence  \cite{TC1,TC2,LiXH}. Nowadays, hyperentanglement has been recognized as a fascinating resource and provided  other important applications, such as linear optical quantum dense coding \cite{dense},  complete photonic Bell-state analysis with linear optics \cite{BSA1,BSA2}, teleportation-based quantum networking \cite{teleportation},  deterministic entanglement purification \cite{Purification1,Purification2,Purification3}, etc. Hyperentangled six-qubit Bell state \cite{hyperentanglement1}, hyperentangled six-qubit cluster state \cite{hyperentanglement2}, and hyperentangled ten-qubit Schr\"{o}dinger cat state \cite{hyperentanglement3} have been experimentally demonstrated in recent years. Quantum teleportation with multiple DOFs \cite{TC1}, complete hyperentangled-Bell- (Greenberger-Horne-Zeilinger-) state analysis \cite{analysis2,analysis3,Error}, hyperparallel quantum repeater \cite{repeater}, hyperentanglement concentration \cite{concentration1,concentration2,concentration3,concentration4,concentration5,concentration6}, and hyperentanglement purification \cite{hyper-purification1,hyper-purification2} have been proposed for high-capacity long distance quantum communication.  Hyperparallel quantum computing also attracted much attention and made a series of outstanding achievements as its promising merits, especially in the field of hyper-parallel universal quantum gates \cite{hypergate1,hypergate2,hypergate3}. Multiple DOFs have been shown potential advantages in simplifying quantum circuits \cite{simplify1,simplify2},  optimizing quantum algorithms \cite{algorithm1}, and improving some traditional strategies \cite{algorithm2}.

Photons are nowadays recognized as excellent candidates for hyperparallel QIP due to their wide range of exploitable DOFs, including spatial \cite{hyperentanglement2}, polarization \cite{hyperentanglement3}, orbital angular momentum \cite{orbital}, transverse \cite{transverse},  frequency \cite{frequency}, spectral \cite{spectral}, time of arrival \cite{time}, etc. Another
showing benefits of photons are high-speed transmission, negligible decoherence, outstanding fast speed, accurate single-qubit operation, and vast photonic industry facilitates. However, a major hurdle for hyperparallel photonic QIP is realizing strongly interactions between individual photons. KLM scheme \cite{KLM}, based on linear optical elements and single photon detectors, is served as a steeping stone for linear undeterministic quantum computing. Currently, photon-mediated (such as cross-Kerr \cite{cross-Kerr1,cross-Kerr2,cross-Kerr3}, neutral atoms \cite{Duan1,netural,atom1,atom2}, atom ensembles \cite{hypergate1,ensemble2}, and artificial atoms \cite{Hu2,QD1,QD2,QD3,NV,superconduct}) interactions are often employed to overcome the intrinsic weak interactions between individual photons charter of parallel and hyperparallel photonic quantum computing. In recent years, artificial atoms (quantum dot in semiconductors \cite{Hu2,QD1,QD2,QD3},  nitrogen vacancy defect centre in diamond \cite{NV}, superconductor \cite{superconduct}) have been received growing interest due to their relatively long coherence time \cite{coherence11,coherence22}, sensitive and quick manipulation, high-fidelity readout \cite{manipulation,readout1,readout2}, custom-designed features \cite{designed,Artificial}, as well as much large linewidths.  Quantum dots (QDs) provide a better matter qubit system \cite{Condition,Fidelity} because they could be designed to have certain characteristics and be assembled in large arrays. Besides, they support microsecond coherence time \cite{coherence11} and picosecond time scale single-qubit rotations \cite{rotation} as well.

Quantum transistor, router, and dynamic random access memory (DRAM) are the key resources for secure quantum network \cite{communication}, metrology \cite{metrology}, and fundamental tests of physics theory \cite{test}. Particularly, quantum transistor provides a potential solution to mitigate transmission loss; quantum router can correct directly the signal from its source to its intended destination conditional on the state of the control qubit; DRAM is characterized by high integration mainly used in large-capacity memory, which can download, store and read out the information. Previous works about these quantum optical elements primarily acted on the single DOF systems \cite{amplification1,amplification2}.  In this paper, we focus on designing compact quantum circuits for implementing  hyperparallel single photon transistor, router and DRAM, respectively. The computing qubits are encoded in the polarization and the spatial DOFs of single photons. The individual photons are bridged by QD mediates confined in  double-sided microcavities. Our schemes have some characters: (1) The inevitable imperfect operations in the QD-cavity unites are taken into account, and the unity fidelities can be achieved in principle. (2) The coupling strength limitations in the microcavities are not necessary. (3) Our presented schemes, different from the traditional ones that operate  on  single DOF, have independent performance on polarization and spatial DOFs, simultaneously. (4) The strong light (source light beam) in quantum transistor is controlled by the weak light (gate photon), and $N$-photon hyper-entanglement state can be generated by means of our transistor.


\section{Hyper-transistor via practice QD-cavity emitter}\label{sec2}

\begin{figure} [htp]   
\centering
\includegraphics[width=6cm,angle=0]{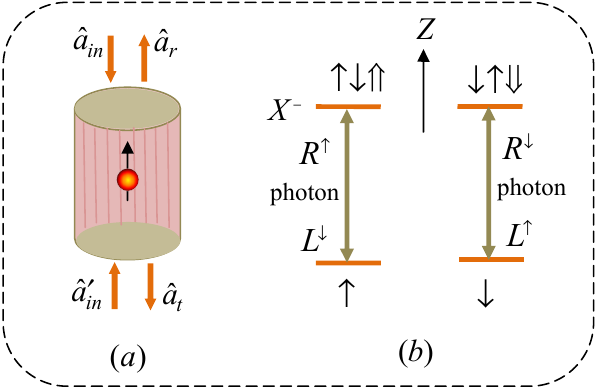}
\caption{(a) A schematic diagram of a singly charged QD confined in a double-sided microcavity. (b) Energy levels and the spin-dependent optical transition rules for a charged QD-cavity emitter. $
|R^\uparrow\rangle$ ($|L^\downarrow\rangle$) represents the propagation direction of the right- (left-) circularly polarized photon is parallel (antiparallel) to the growth axis of the QD.
$|\Uparrow\rangle$ and $|\Downarrow\rangle$ denote the heavy-hole spin states $|\pm3/2\rangle$, respectively. $|\uparrow\rangle$ and $|\downarrow\rangle$ indicate the electron spin states
$|\pm1/2\rangle$, respectively.}\label{Level}
\end{figure}

The key ingredient of the optical-QD-based QIP is the realization of entanglement between a QD spin and a single photon. In 2009,  Hu \emph{et al.} \cite{Hu2} proposed a QD-microcavity emitter,  i.e., a singly charged QD [e.g., a self-assembled Al(Ga)As QD or GaAs QD] placed in the center of a double-sided optical microcavity. As shown in Fig. \ref{Level}, a negatively charged exciton $X^-$ consists of two electrons bound to one heavy hole \cite{IC}. The ground states and the excited states of this singly charged QD are the electron spin states and the exciton $X^-$ spin states, respectively \cite{Hu2}. 

The $X^-$ exhibits spin-dependent optical transition rules due to the Pauli's exclusion rules and the conservation of total spin angular momentum \cite{PL}.  In detail, if the electron is in the state $|\uparrow\rangle$, only the circularly polarized photon with $S_z=+1$ (marked by $|L^{\downarrow}\rangle$ or $|R^{\uparrow}\rangle$) feels the ``hot" cavity and couples to the transition $|\uparrow\rangle\leftrightarrow|\uparrow\downarrow\Uparrow\rangle$. If the electron is in the state $|\downarrow\rangle$, only the circularly polarized light with $S_z=-1$ (marked by $|L^{\uparrow}\rangle$ or $|R^{\downarrow}\rangle$) feels the ``hot" cavity and couples to the transition $|\downarrow\rangle\leftrightarrow|\downarrow\uparrow\Downarrow\rangle$. Here, $|R^\uparrow\rangle$  and $|L^\uparrow\rangle$ ($|R^\downarrow\rangle$ and $|L^\downarrow\rangle$) denote the propagation direction of the right- and left- circularly polarized photon is parallel (antiparallel) to the spin quantization axis ($z$ axis). $|\uparrow\rangle$ and $|\downarrow\rangle$ are the electron spin states with $J=\pm1/2$, respectively. $|\Uparrow\rangle$ and $|\Downarrow\rangle$ are the heavy hole spin states with $J=\pm3/2$, respectively.

The reflection/transmission coefficients of the hot/cold cavities can be obtained by solving Heisenberg equations of motion for the cavity field operator $\hat{a}$ and the dipole operator $\hat{\sigma}_-$ \cite{Heisenberg} and the input-output relations between the input and output fields

\begin{eqnarray}          \label{eq1}
&&\frac{d\hat{a}}{dt}=-\left[i(\omega_c-\omega)+\kappa+\frac{\kappa_s}{2}\right]\hat{a}-\text{g}\;\hat{\sigma}_{-}-\sqrt{\kappa}\,\hat{a}_{in}-\sqrt{\kappa}\,\hat{a}_{in}'+\hat{H}, \nonumber\\
&&\frac{d\hat{\sigma}_-}{dt}=-\left[i(\omega_{X^-}-\omega)+\frac{\gamma}{2}\right]\hat{\sigma}_{-}-\text{g}\;\sigma_z\;\hat{a}+\hat{G}, \nonumber\\
&&\hat{a}_r =\hat{a}_{in}+ \sqrt{\kappa}\,\hat{a},  \nonumber\\
&&\hat{a}_t=\hat{a}_{in}'+\sqrt{\kappa}\,\hat{a}.
\end{eqnarray}
Here, $\omega$, $\omega_c$, and $\omega_{X^-}$ are the frequencies of the single photon, the cavity mode, and the $X^-$ dipole transition, respectively. $\text{g}$ is the coupling strength of the cavity-$X^-$ combination. $\gamma/2$, $\kappa$, and $\kappa_s/2$ are the decay rates of the $X^-$ dipole, the cavity field, and the side leakage, respectively. $\hat{a}_{in}$ and $\hat{a}_{in}'$ ($\hat{a}_r$ and $\hat{a}_t$) are the cavity input (output) fields. $\sigma_z$ is the inversion operator of the singly charged QD. $\hat{H}$ and $\hat{G}$ are the noise operators.

When $X^-$ predominantly stays in the ground states, i.e., taking $\langle\sigma_z\rangle\approx-1$, the reflection coefficient $r(\omega)$ and the transmission coefficient $t(\omega)$ of the QD-microcavity system can be written as \cite{Hu2,coefficients},
\begin{eqnarray}             \label{eq2}
&&r(\omega)=1+t(\omega), \nonumber\\
&&t(\omega)=\frac{-\kappa\left[i(\omega_{X^{-}}-\omega)+\frac{\gamma}{2}\right]}{\left[i(\omega_{X^{-}}-\omega)+\frac{\gamma}{2}\right]\left[i(\omega_c-\omega)+\kappa+\frac{\kappa_s}{2}\right]+\text{g}^2}.
\end{eqnarray}
%
%
In the practice experiment, the inevitable imperfect birefringence of the cavity will reduce the fidelity and the efficiency of the emitter by a few percents. Therefore, the spin-dependent transition rules can be summarized as
\begin{eqnarray}                  \label{eq3}
&&|R^\uparrow   \uparrow\rangle \rightarrow  r|L^\downarrow \uparrow\rangle  +t|R^\uparrow \uparrow\rangle,\quad
|R^\uparrow   \downarrow\rangle \rightarrow t_0|R^\uparrow \downarrow\rangle + r_0|L^\downarrow \downarrow\rangle, \nonumber\\\;
&&|L^\downarrow \uparrow\rangle \rightarrow  r|R^\uparrow \uparrow\rangle  +t|L^\downarrow \uparrow\rangle,\quad
|L^\downarrow\downarrow\rangle \rightarrow t_0|L^\downarrow\downarrow\rangle + r_0|R^\uparrow\downarrow\rangle, \nonumber\\
&&|L^\uparrow  \downarrow\rangle \rightarrow r|R^\downarrow  \downarrow\rangle +t|L^\uparrow,\downarrow\rangle,\quad
|R^\downarrow \uparrow\rangle \rightarrow t_0|R^\downarrow \uparrow\rangle  +r_0|L^\uparrow \uparrow\rangle, \nonumber\\
&&|R^\downarrow \downarrow\rangle \rightarrow r|L^\uparrow \downarrow\rangle +t|R^\downarrow \downarrow\rangle, \quad
|L^\uparrow   \uparrow\rangle \rightarrow t_0|L^\uparrow \uparrow\rangle  +r_0|R^\downarrow \uparrow\rangle.
\end{eqnarray}
Here $r_0$ and $t_0$ are described by Eq. (\ref{eq2}) with $\text{g}=0$. If side leakage is not taken into account (i.e. $\kappa_s=0$) and $\text{g}\gg2 \gamma\kappa$, and then $t=0$, $t_0=-1$, $r=1$, and $r_0=0$. In experiment, such ideal conditional is a challenge.


The spin-dependent Kerr nonlinearity shown in Eq. (\ref{eq3}) can be used to implement hyper-parallel photonic elements in the following sections. We design compact quantum circuits for implementing hyperparallel transistor (hyper-transistor), hyperparallel router (hyper-router)  and hyperparallel DRAM (hyper-DRAM) encoded in the polarization and the spatial DOFs in the single-photon systems.


\subsection{P-transistor via practice QD-cavity emitter}\label{sec2-1}

\begin{figure} [htp]
\centering
\includegraphics[width=11.5 cm,angle=0]{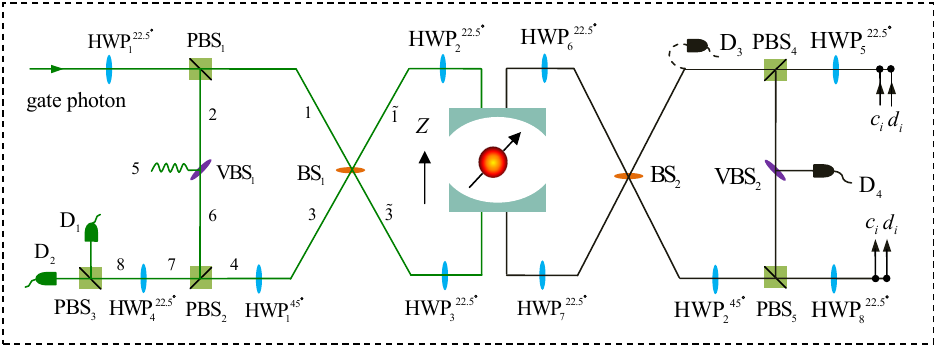}
\caption{A description for implementing a p-transistor.
HWP$^{22.5^\circ}_{1,\cdots,8}$ with  using  half wave plates rotated at 22.5$^\circ $ represent Hadamard operations on polarization DOF.
HWP$^{45^\circ}_{1,2}$ stand for half wave plate oriented at 45$^\circ $ performing bit-flip operations $\sigma_{p,x}=|R\rangle\langle L|+|L\rangle\langle R|$.
PBS$_{1,\cdots,4}$ are circularly polarizing beam splitters.
BS$_{1,2}$ are nonpolarizing balanced beam splitters performing Hadamard operations on the spatial DOF, i.e., $|l^1\rangle\leftrightarrow(|l^{\tilde{1}}\rangle+|l^{\tilde{3}}\rangle)/\sqrt{2}$, $|l^3\rangle\leftrightarrow(|l^{\tilde{1}}\rangle-|l^{\tilde{3}}\rangle)/\sqrt{2}$.
VBS$_{1,2}$ are adjustable beam splitters with transmission coefficient ($t-t_0$) and reflection coefficient $(\sqrt{1-(t-t_0)^2})$.
$D_i$ ($i=1,\cdots, 4$) are single-photon detectors.} \label{Transistor-polarization}
\end{figure}

The hyper-transistor amplifies both an arbitrary polarization state to the same state encoded on $N$ photons (p-transistor)  and an arbitrary spatial state to the same state encoded on $N$ photons (s-transistor), simultaneously. The framework of our p-transistor without effecting the spatial DOF is shown in Fig. \ref{Transistor-polarization}. Suppose that the states of the gate photon and the QD spin are initially prepared as
\begin{eqnarray}                  \label{eq4}
&&|\psi\rangle_{\text{gate photon}}=\alpha|R\rangle+\beta|L\rangle, \nonumber\\
&&|\psi\rangle_{\text{electron}}=\frac{1}{\sqrt{2}}(|\uparrow\rangle-|\downarrow\rangle),
\end{eqnarray}
where $\alpha$ and $\beta$ are the arbitrary complex numbers and satisfy $|\alpha|^2+|\beta|^2=1$.

First, the gate photon is injected. As shown in Fig. \ref{Transistor-polarization}, before and after the gate photon passes through the building block composed of PBS$_1$, PBS$_2$, BS$_1$, VBS$_1$, HWP$^{22.5^\circ}_2$, HWP$^{22.5^\circ}_3$, QD, and HWP$^{45^\circ}_1$, Hadamard operations are performed on it by using HWP$^{22.5^\circ}_1$  and HWP$^{22.5^\circ}_4$, respectively. Specifically, circularly polarizing beam splitters, PBS$_{1,2}$, transmit the $R$-polarized wave packets and reflect the $L$-polarized wave packets, respectively. HWP$^{45^\circ}_1$ denotes a half-wave plate aligned at 45$^\circ$ to complete the bit-flip operation $\sigma_{p,x}=|R\rangle\langle L|+|L\rangle\langle R|$ on the passing photons. HWP$^{22.5^\circ}_{1,\cdots,4}$ are half-wave plates oriented at 22.5$^\circ$ to achieve polarization Hadamard operations
\begin{eqnarray}                  \label{eq5}
|R\rangle\leftrightarrow\frac{1}{\sqrt{2}}(|R\rangle+|L\rangle),\qquad
|L\rangle\leftrightarrow\frac{1}{\sqrt{2}}(|R\rangle-|L\rangle).
\end{eqnarray}
The nonpolarizing balanced beam splitter, BS$_1$, induces a Hadamard operation on the spatial states $|l^1\rangle$, $|l^{\tilde{1}}\rangle$, $|l^3\rangle$, and  $|l^{\tilde{3}}\rangle$ (spatial Hadamard operation)
\begin{eqnarray}                  \label{eq6}
&&|l^1\rangle\leftrightarrow\frac{1}{\sqrt{2}}(|l^{\tilde{1}}\rangle+|l^{\tilde{3}}\rangle), \quad
|l^3\rangle\leftrightarrow\frac{1}{\sqrt{2}}(|l^{\tilde{1}}\rangle-|l^{\tilde{3}}\rangle), \nonumber\\
&&|l^{\tilde{1}}\rangle\leftrightarrow\frac{1}{\sqrt{2}}(|l^{1}\rangle+|l^{3}\rangle), \quad
|l^{\tilde{3}}\rangle\leftrightarrow\frac{1}{\sqrt{2}}(|l^{1}\rangle-|l^{3}\rangle).
\end{eqnarray}
The adjustable beam splitter, VBS$_1$, has a variable transmission coefficient $t-t_0$ and reflection coefficient $\sqrt{1-(t-t_0)^2}$, and it can be achieved by using two 50:50 BSs and two phase shifters \cite{VBS}.
Therefore, operations
($\text{HWP}^{22.5^\circ}_1\rightarrow \text{PBS}_1\rightarrow \text{BS}_1\rightarrow \text{HWP}^{22.5^\circ}_{2,3}
\rightarrow \text{QD}\rightarrow \text{HWP}^{22.5^\circ}_{2,3}\rightarrow \text{BS}_1\rightarrow\text{HWP}^{45^\circ}_{1}$) transform the state of the gate photon together with QD from $|\psi_p\rangle_0$ to $|\psi_p\rangle_1$. Here
\begin{eqnarray}                  \label{eq7}
|\psi_p\rangle_0=\frac{1}{\sqrt{2}}(\alpha|R\rangle+\beta|L\rangle)\otimes(|\uparrow\rangle-|\downarrow\rangle),
\end{eqnarray}
\begin{eqnarray}                  \label{eq8}
|\psi_p\rangle_1&=&\frac{1}{2}[\alpha(r+t_0)|R^{1}\uparrow\rangle+\alpha(t-t_0)|R^{4}\uparrow\rangle+\beta(r+t_0)|R^{1}\uparrow\rangle+\beta(t-t_0)|R^{4}\uparrow\rangle \nonumber\\&&
-\alpha(r+t_0)|R^{1}\downarrow\rangle+\alpha(t-t_0)|R^{4}\downarrow\rangle-\beta(r+t_0)|R^{1}\downarrow\rangle+\beta(t-t_0)|R^{4}\downarrow\rangle \nonumber\\
&&+\alpha|L^2\uparrow\rangle-\beta|L^2\uparrow\rangle-\alpha|L^2\downarrow\rangle+\beta|L^2\downarrow\rangle].
\end{eqnarray}
$|R^i\rangle$ ($|L^i\rangle$) represents the $R$-polarized ($L$-polarized) photon emitted from the spatial mode $i$.

Second, after the wave packet $|L^2\rangle$ interacts with VBS$_1$, the total state of the system is evolved to
\begin{eqnarray}                  \label{eq9}
|\psi_p\rangle_2&=&\frac{1}{2}[\alpha(t-t_0)|R^{4}\uparrow\rangle+\beta(t-t_0)|R^{4}\uparrow\rangle+\alpha(t-t_0)|R^{4}\downarrow\rangle+\beta(t-t_0)|R^{4}\downarrow\rangle
 \nonumber\\&&
+\alpha(t-t_0)|L^{6}\uparrow\rangle-\beta(t-t_0)|L^{6}\uparrow\rangle-\alpha(t-t_0)|L^{6}\downarrow\rangle+\beta(t-t_0)|L^{6}\downarrow\rangle
 \nonumber\\&&
+\alpha(r+t_0)|R^{1}\uparrow\rangle+\beta(r+t_0)|R^{1}\uparrow\rangle-\alpha(r+t_0)|R^{1}\downarrow\rangle-\beta(r+t_0)|R^{1}\downarrow\rangle
 \nonumber\\&&
+(\sqrt{1-(t-t_0)^2})(\alpha|L^{5}\uparrow\rangle-\beta|L^{5}\uparrow\rangle-\alpha|L^{5}\downarrow\rangle+\beta|L^{5}\downarrow\rangle)].
\end{eqnarray}

Third, the wave packets $|R^4\rangle$ and $|L^6\rangle$ arrive at PBS$_2$, simultaneously, and PBS$_2$ makes $|\psi_p\rangle_2$ into
\begin{eqnarray}                  \label{eq10}
|\psi_p\rangle_3&=&\frac{1}{2}[\alpha(t-t_0)|R^{7}\uparrow\rangle+\beta(t-t_0)|R^{7}\uparrow\rangle+\alpha(t-t_0)|R^{7}\downarrow\rangle+\beta(t-t_0)|R^{7}\downarrow\rangle
 \nonumber\\&&
+\alpha(t-t_0)|L^{7}\uparrow\rangle-\beta(t-t_0)|L^{7}\uparrow\rangle-\alpha(t-t_0)|L^{7}\downarrow\rangle+\beta(t-t_0)|L^{7}\downarrow\rangle
 \nonumber\\&&
+\alpha(r+t_0)|R^{1}\uparrow\rangle+\beta(r+t_0)|R^{1}\uparrow\rangle-\alpha(r+t_0)|R^{1}\downarrow\rangle-\beta(r+t_0)|R^{1}\downarrow\rangle
 \nonumber\\&&
+(\sqrt{1-(t-t_0)^2})(\alpha|L^{5}\uparrow\rangle-\beta|L^{5}\uparrow\rangle-\alpha|L^{5}\downarrow\rangle+\beta|L^{5}\downarrow\rangle)].
\end{eqnarray}

Fourth, the gate photon emitted from the spatial mode $7$ is detected in the basis $\{(|R\rangle\pm|L\rangle)/\sqrt{2}\}$ by HWP$^{22.5^\circ}_4$ and single photon detectors $D_1$ and $D_2$. In detail, on detecting the gate photon in $(|R\rangle-|L\rangle)/\sqrt{2}$, we project $|\psi_p\rangle_3$ into
\begin{eqnarray}                  \label{eq11}
|\psi_p\rangle_4=\alpha|\downarrow\rangle+\beta|\uparrow\rangle.
\end{eqnarray}
Next, we inject a single photon, or an ultrafast ps or fs  $(\pi)_y$ optical pulse  to perform bit-flip operation $\sigma_{e,x}=|\uparrow\rangle\langle\downarrow|+|\downarrow\rangle\langle\uparrow|$ on the QD spin \cite{rotation}
to obtain the desired state
\begin{eqnarray}                  \label{eq12}
|\psi_p'\rangle_4=\alpha|\uparrow\rangle+\beta|\downarrow\rangle.
\end{eqnarray}
Alternatively, on detecting the gate photon in $(|R\rangle+|L\rangle)/\sqrt{2}$, a desired state described by Eq. (\ref{eq12}) is obtained directly.

Fifth, source photon 1 in the normalization state $|R_1\rangle(\zeta_1|c_1\rangle+\xi_1|d_1\rangle)$ is injected into the building block consisted of HWP$^{22.5^\circ}_5$, PBS$_4$, BS$_2$, HWP$^{22.5^\circ}_6$, HWP$^{22.5^\circ}_7$, QD, VBS$_2$, $D_3$, $D_4$, HWP$^{45^\circ}_2$, PBS$_5$, and HWP$^{22.5^\circ}_8$. If $D_3$ and $D_4$ do not click, the state of the system will collapse into
\begin{eqnarray}                  \label{eq13}
|\psi_p\rangle_5=(t-t_0)(\alpha|R_1\uparrow\rangle-\beta|L_1\downarrow\rangle)\otimes(\zeta_1|c_1\rangle+\xi_1|d_1\rangle).
\end{eqnarray}
Here $c_i$ and  $d_i$ are the two spatial modes of the source photon $i$.
Otherwise, the system is projected into the spin state described by Eq. (\ref{eq12}) (i.e., $\alpha|\uparrow\rangle+\beta|\downarrow\rangle$).
That is, the scheme is fail, and then we need to repeat above arguments.

Sixth, repeating above process from source photon 2 to $N$ in succession, after the source photons interact with the QD, if $D_3$ and $D_4$ are not clicked,
the joint state collapses into
\begin{eqnarray}                  \label{eq14}
|\psi_p\rangle_6&=&(t-t_0)^N[\alpha|R_1R_2 \cdots R_N\uparrow\rangle+(-1)^N\beta|L_1L_2 \cdots L_N\downarrow\rangle] \nonumber\\&&\otimes(\zeta_1|c_1\rangle+\xi_1|d_1\rangle)\otimes(\zeta_2|c_2\rangle+\xi_2|d_2\rangle)\otimes \cdots \otimes(\zeta_N|c_N\rangle+\xi_N|d_N\rangle).
\end{eqnarray}
Here, the different spatial modes, i.e., $c_i$ and $d_i$, of the source photon $i$ can be separated by spatial with optical switch \cite{Switch1,Switch2}, or time.

Seventh, to complete the p-transistor, we measure the spin of the QD in the basis $\{|\pm\rangle=(|\uparrow\rangle\pm|\downarrow\rangle)/\sqrt{2}\}$ and apply some feed-forward operations on one of the outing photons to obtain the desired state
\begin{eqnarray}                  \label{eq15}
|\psi_p\rangle_7&=&(t-t_0)^N(\alpha|R_1R_2 \cdots R_N\rangle+\beta|L_1L_2 \cdots L_N\rangle) \nonumber\\&&\otimes(\zeta_1|c_1\rangle+\xi_1|d_1\rangle)\otimes(\zeta_2|c_2\rangle+\xi_2|d_2\rangle)\otimes \cdots \otimes(\zeta_N|c_N\rangle+\xi_N|d_N\rangle).
\end{eqnarray}
In detail, on detecting the QD spin in $|+\rangle$, if and only if $N$ is odd number, we apply a single-qubit operation $\sigma_{p,z}=|R\rangle\langle R|-|L\rangle\langle L|$ on one of the outing photons. On detecting the QD spin in $|-\rangle$, if and only if $N$ is even number, we apply a single-qubit operation $\sigma_{p,z}$  on one of the outing photons.


\subsection{S-transistor via practice QD-cavity emitter}\label{sec2-2}

\begin{figure} [htp]
\centering
\includegraphics[width=9.0 cm,angle=0]{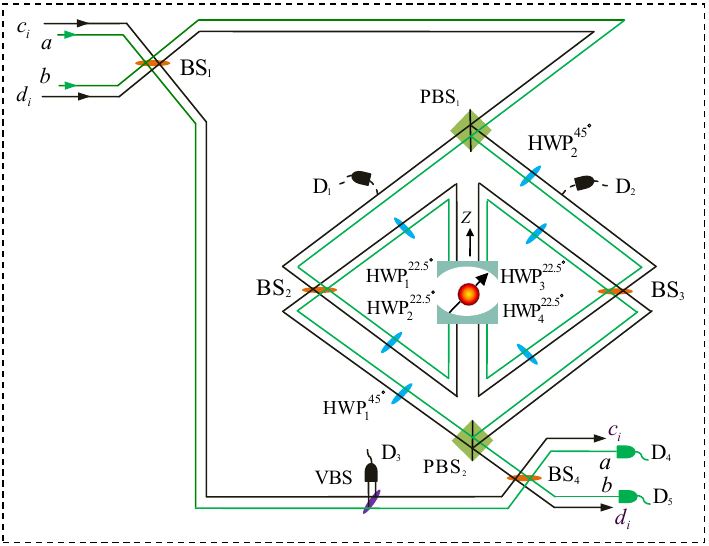}
\caption{ A schematic diagram for implementing a s-transistor. } \label{Transistor-spatial}
\end{figure}

Up to now, p-transistor has been completed. However, in order to implement a single-photon transistor performing on the polarization and spatial DOFs, simultaneously, a s-transistor should be designed in this subsection (see Fig. \ref{Transistor-spatial}).

First, the gate photon in the state $(\alpha|R\rangle+\beta|L\rangle)\otimes(\gamma|a\rangle+\delta|b\rangle)$ is injected and arrives at BS$_1$. BS$_1$ induces the state of gate photon and the QD from
\begin{eqnarray}                  \label{eq16}
|\psi_s\rangle_0=\frac{1}{\sqrt{2}}(\alpha|R\rangle+\beta|L\rangle)\otimes(\gamma|a\rangle+\delta|b\rangle)\otimes(|\uparrow\rangle-|\downarrow\rangle),
\end{eqnarray}
into
\begin{eqnarray}                  \label{eq17}
|\psi_s\rangle_1=\frac{1}{2}(\alpha|R\rangle+\beta|L\rangle)(\gamma|a\rangle+\gamma|b\rangle+\delta|a\rangle-\delta|b\rangle)(|\uparrow\rangle-|\downarrow\rangle).
\end{eqnarray}
Here, $a$ and  $b$ are the two spatial modes of the gate photon for implementing s-transistor, and $|\alpha|^2+|\beta|^2=1$, $|\gamma|^2+|\delta|^2=1$.

Second, the wave packets emitted from the spatial mode $b$ arrive at VBS directly. Alternately, the $R$-polarized ($L$-polarized) component emitted from the spatial mode $a$ passes through the building block comprised of
PBS$_1$, BS$_2$, HWP$^{22.5^\circ}_1$, QD, HWP$^{22.5^\circ}_2$, HWP$^{45^\circ}_1$, and PBS$_2$
(PBS$_1$, HWP$^{45^\circ}_2$, BS$_3$, HWP$^{22.5^\circ}_3$, QD, HWP$^{22.5^\circ}_4$, and PBS$_2$).
These operations transform $|\psi_s\rangle_1$ into
\begin{eqnarray}                  \label{eq18}
|\psi_s\rangle_2&=&\frac{1}{2}(t-t_0)(\alpha|R^{l}\uparrow\rangle+\alpha|R^{l}\downarrow\rangle+\beta|L^{r}\uparrow\rangle+\beta|L^{r}\downarrow\rangle)(\gamma|a\rangle+\delta|a\rangle) \nonumber\\&&
+\frac{1}{2}(t-t_0)(\alpha|R\rangle+\beta|L\rangle)(|\uparrow\rangle-|\downarrow\rangle)(\gamma|b\rangle-\delta|b\rangle) \nonumber\\&&
+\frac{1}{2}(r+t_0)(\alpha|R^{l,D_1}\uparrow\rangle-\alpha|R^{l,D_1}\downarrow\rangle+\beta|R^{r,D_2}\uparrow\rangle-\beta|R^{r,D_2}\downarrow\rangle)(\gamma|a\rangle+\delta|a\rangle) \nonumber\\&&
+\frac{1}{2}\sqrt{1-(t-t_0)^2}(\alpha|R^{D_3}\rangle+\beta|L^{D_3}\rangle)(|\uparrow\rangle-|\downarrow\rangle)(\gamma|b\rangle-\delta|b\rangle).
\end{eqnarray}
Here, the superscript $l$ ($r$) denotes the wave packets emitted from the left (right) arm, and the superscript $D_1$ ($D_2$) denotes the wave packets will be detected by detector $D_1$ ($D_2$).
PBS$_2$ transforms $|\psi_s\rangle_2$ into
\begin{eqnarray}                  \label{eq19}
|\psi_s\rangle_3&=&\frac{1}{2}(t-t_0)(\alpha|R^{r}\uparrow\rangle+\alpha|R^{r}\downarrow\rangle+\beta|L^{r}\uparrow\rangle+\beta|L^{r}\downarrow\rangle)(\gamma|a\rangle+\delta|a\rangle) \nonumber\\&&
+\frac{1}{2}(t-t_0)(\alpha|R\rangle+\beta|L\rangle)(|\uparrow\rangle-|\downarrow\rangle)(\gamma|b\rangle-\delta|b\rangle) \nonumber\\&&
+\frac{1}{2}(r+t_0)(\alpha|R^{l,D_1}\uparrow\rangle-\alpha|R^{l,D_1}\downarrow\rangle+\beta|R^{r,D_2}\uparrow\rangle-\beta|R^{r,D_2}\downarrow\rangle)(\gamma|a\rangle+\delta|a\rangle) \nonumber\\&&
+\frac{1}{2}\sqrt{1-(t-t_0)^2}(\alpha|R^{D_3}\rangle+\beta|L^{D_3}\rangle)(|\uparrow\rangle-|\downarrow\rangle)(\gamma|b\rangle-\delta|b\rangle).
\end{eqnarray}

Third, as shown in Fig. \ref{Transistor-spatial},  after the wave packets emitted form the right arm and the spatial mode $b$ mix at BS$_4$, and then $|\psi_s\rangle_3$ is evolved as
\begin{eqnarray}                  \label{eq20}
|\psi_s\rangle_4&=&\frac{1}{\sqrt{2}}(t-t_0)[|a\rangle(\gamma|\uparrow\rangle+\delta|\downarrow\rangle)+|b\rangle(\gamma|\downarrow\rangle+\delta|\uparrow\rangle)]\otimes(\alpha|R\rangle+\beta|L\rangle) \nonumber\\&&
+\frac{1}{2}(r+t_0)(\alpha|R^{l,D_1}\uparrow\rangle-\alpha|R^{l,D_1}\downarrow\rangle+\beta|R^{r,D_2}\uparrow\rangle-\beta|R^{r,D_2}\downarrow\rangle)(\gamma|a\rangle+\delta|a\rangle) \nonumber\\&&
+\frac{1}{2}\sqrt{1-(t-t_0)^2}(\alpha|R^{D_3}\rangle+\beta|L^{D_3}\rangle)(|\uparrow\rangle-|\downarrow\rangle)(\gamma|b\rangle-\delta|b\rangle).
\end{eqnarray}

Fourth, the spatial modes of the outing photon are measured. In detail, on detecting the gate photon in the spatial mode $a$ and $D_1$, $D_2$ and $D_3$ do not click, we project $|\psi_s\rangle_4$ into the desired state
\begin{eqnarray}                  \label{eq21}
|\psi_s\rangle_5=(\gamma|\uparrow\rangle+\delta|\downarrow\rangle)\otimes(\alpha|R\rangle+\beta|L\rangle).
\end{eqnarray}
Alternatively, on detecting the gate photon in the spatial mode $b$ and $D_1$, $D_2$ and $D_3$ do not click, we project $|\psi_s\rangle_4$ into the state
\begin{eqnarray}                  \label{eq22}
|\psi_s'\rangle_5=(\gamma|\downarrow\rangle+\delta|\uparrow\rangle)\otimes(\alpha|R\rangle+\beta|L\rangle).
\end{eqnarray}
And then,  we perform a bit-flip operation $\sigma_x$ on the QD spin to obtain the desired state described by Eq. (\ref{eq21}).

Fifth, repeating above process from photon 1 to $N$. After the $N$ photons in the state $|c_i\rangle(\alpha_i|R\rangle+\beta_i|L\rangle)$ pass through the block in succession,  when $D_1$, $D_2$ and $D_3$ are not clicked, the joint state collapses into
\begin{eqnarray}                  \label{eq23}
|\psi_s\rangle_6&=&(t-t_0)^N[\gamma |c_1c_2...c_N\rangle|\uparrow\rangle+(-1)^N\delta|d_1d_2...d_N\rangle|\downarrow\rangle)] \nonumber\\&&
\otimes(\alpha_1|R\rangle+\beta_1|L\rangle)\otimes(\alpha_2|R\rangle+\beta_2|L\rangle)\otimes...\otimes(\alpha_N|R\rangle+\beta_N|L\rangle).
\end{eqnarray}

Sixth, we measure the spins of the QD in the basis $\{|\pm\rangle\}$ and apply some proper feed-forward operations on the spatial DOF to
complete the s-transistor, i.e., to achieve the state
\begin{eqnarray}                  \label{eq24}
|\psi_s\rangle_7&=&(t-t_0)^N(\gamma |c_1c_2...c_N\rangle+\delta|d_1d_2...d_N\rangle) \nonumber\\&&
\otimes(\alpha_1|R\rangle+\beta_1|L\rangle)\otimes(\alpha_2|R\rangle+\beta_2|L\rangle)\otimes...\otimes(\alpha_N|R\rangle+\beta_N|L\rangle).
\end{eqnarray}
In detail, on detecting the electronic state $|+\rangle$, if and only if $N$ is odd, we apply a single-qubit operation $\sigma_z$ on one of the spatial modes to correct the minus sign. On detecting the electronic state $|-\rangle$, if and only if $N$ is even, we apply a single-qubit operation $\sigma_z$  on one of the spatial modes.


\section{Hyper-router via practice QD-cavity emitter}\label{sec3}

Quantum router \cite{router} is the key quantum technology for quantum networks and quantum computers. It directs a signal qubit to its intended destination according to the state of the control qubits, but keeping the state of signal qubit is unchanged. In this section, let us introduce the action of our hyper-router acting on polarization and spatial DOFs, simultaneously.

\begin{figure} [htp]
\centering
\includegraphics[width=7.5 cm,angle=0]{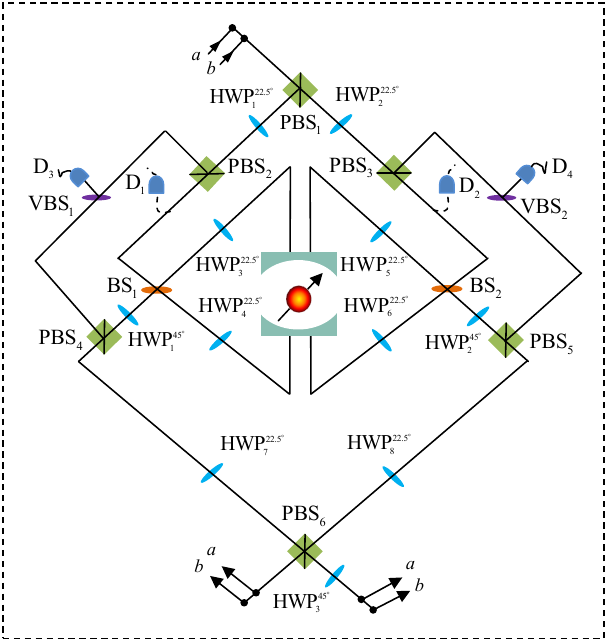}
\caption{ Schematic diagram of hyper-router.}
\label{Router-polarization}
\end{figure}

As shown in Fig. \ref{Router-polarization}, the photon is used as the signal qubit in the normalization state $(\alpha|R\rangle+\beta|L\rangle)\otimes(\delta_1|a\rangle+\delta_2|b\rangle)
$, the QD spin is served  as the control qubit in the normalization state $(\gamma|\uparrow\rangle+\eta|\downarrow\rangle)$.

First, the photon is injected and followed by PBS$_1$, HWP$^{22.5^\circ}_1$  (HWP$^{22.5^\circ}_2$). Operations ($\rm PBS_1 \rightarrow HWP^{22.5^\circ}_1$ and $\rm PBS_1 \rightarrow HWP^{22.5^\circ}_2)$ transform the system from the initialization state $|\varphi\rangle_0$ to $|\varphi\rangle_1$. Here
\begin{eqnarray}                  \label{eq25}
|\varphi\rangle_0=(\alpha|R\rangle+\beta|L\rangle)\otimes(\delta_1|a\rangle+\delta_2|b\rangle)\otimes(\gamma|\uparrow\rangle+\eta|\downarrow\rangle), \end{eqnarray}
\begin{eqnarray}                  \label{eq26}
|\varphi\rangle_1=\frac{1}{\sqrt{2}}(\alpha(|R^r\rangle+|L^r\rangle)+\beta(|R^l\rangle-|L^l\rangle))\otimes(\delta_1|a\rangle+\delta_2|b\rangle)\otimes(\gamma|\uparrow\rangle+\eta|\downarrow\rangle).
\end{eqnarray}

Second, the wave packets emitted from the left round $|R^l\rangle$ and $|L^l\rangle$  (the right round $|R^r\rangle$ and $|L^r\rangle$) pass through the block composed of PBS$_2$, VBS$_1$, BS$_1$, QD, HWP$^{22.5^\circ}_3$, HWP$^{22.5^\circ}_4$, HWP$^{45^\circ}_1$, and PBS$_4$ (PBS$_3$, VBS$_2$, BS$_2$, QD, HWP$^{22.5^\circ}_5$, HWP$^{22.5^\circ}_6$, HWP$^{45^\circ}_2$, and PBS$_5$).
Such two blocks make $|\varphi\rangle_1$ become
\begin{eqnarray}                  \label{eq27}
|\varphi\rangle_2&=&\frac{1}{\sqrt{2}}(\alpha(t-t_0)|R^r\rangle(\gamma|\uparrow\rangle-\eta|\downarrow\rangle)+\alpha(t-t_0)|L^r\rangle(\gamma|\uparrow\rangle+\eta|\downarrow\rangle) \nonumber\\&&
+\beta(t-t_0)|R^l\rangle(\gamma|\uparrow\rangle-\eta|\downarrow\rangle)-\beta(t-t_0)|L^l\rangle(\gamma|\uparrow\rangle+\eta|\downarrow\rangle))\otimes(\delta_1|a\rangle+\delta_2|b\rangle) \nonumber\\&&
+\frac{1}{\sqrt{2}}(\alpha(r+t_0)|R^{r,D_2}\rangle+\beta(r+t_0)|R^{l,D_1}\rangle+\alpha\sqrt{1-(t-t_0)^2}|L^{r,D_4}\rangle \nonumber\\&&-\beta\sqrt{1-(t-t_0)^2}|L^{l,D_3}\rangle)
\otimes(\gamma|\uparrow\rangle+\eta|\downarrow\rangle)\otimes(\delta_1|a\rangle+\delta_2|b\rangle).
\end{eqnarray}

Third, before the wave packets converge at  PBS$_6$, two Hadamard operations are applied on them by using HWP$^{22.5^\circ}_7$ and HWP$^{22.5^\circ}_8$. That is, HWP$^{22.5^\circ}_7$ and HWP$^{22.5^\circ}_8$ make the system from $|\varphi\rangle_2$ to
\begin{eqnarray}                  \label{eq28}
|\varphi\rangle_3&=&(\alpha(t-t_0)(\gamma|R^r\rangle|\uparrow\rangle-\eta|L^r\rangle|\downarrow\rangle)
+\beta(t-t_0)(\gamma|L^l\rangle|\uparrow\rangle-\eta|R^l\rangle|\downarrow\rangle)) \nonumber\\&&\otimes(\delta_1|a\rangle+\delta_2|b\rangle)
+\frac{1}{\sqrt{2}}(\alpha(r+t_0)|R^{r,D_2}\rangle+\beta(r+t_0)|R^{l,D_1}\rangle \nonumber\\&&+\alpha\sqrt{1-(t-t_0)^2}|L^{r,D_4}\rangle-\beta\sqrt{1-(t-t_0)^2}|L^{l,D_3}\rangle)
 \nonumber\\&&\otimes(\gamma|\uparrow\rangle+\eta|\downarrow\rangle)\otimes(\delta_1|a\rangle+\delta_2|b\rangle).
\end{eqnarray}

Fourth, after PBS$_6$, $|\varphi\rangle_3$ changes into
\begin{eqnarray}                  \label{eq29}
|\varphi\rangle_4&=&(\alpha(t-t_0)(\gamma|R^l\rangle|\uparrow\rangle-\eta|L^r\rangle|\downarrow\rangle)
+\beta(t-t_0)(\gamma|L^l\rangle|\uparrow\rangle-\eta|R^r\rangle|\downarrow\rangle)) \nonumber\\&&\otimes(\delta_1|a\rangle+\delta_2|b\rangle)
+\frac{1}{\sqrt{2}}(\alpha(r+t_0)|R^{r,D_2}\rangle+\beta(r+t_0)|R^{l,D_1}\rangle \nonumber\\&&+\alpha\sqrt{1-(t-t_0)^2}|L^{r,D_4}\rangle-\beta\sqrt{1-(t-t_0)^2}|L^{l,D_3}\rangle)
 \nonumber\\&&\otimes(\gamma|\uparrow\rangle+\eta|\downarrow\rangle)\otimes(\delta_1|a\rangle+\delta_2|b\rangle).
\end{eqnarray}
HWP$^{45^\circ}_3$ flips state of the the outing photon emitted from the right hand, that is, Eq. (\ref{eq29}) becomes
\begin{eqnarray}                  \label{eq30}
|\varphi\rangle_5&=&(\alpha(t-t_0)(\gamma|R^l\rangle|\uparrow\rangle-\eta|R^r\rangle|\downarrow\rangle)
+\beta(t-t_0)(\gamma|L^l\rangle|\uparrow\rangle-\eta|L^r\rangle|\downarrow\rangle)) \nonumber\\&&\otimes(\delta_1|a\rangle+\delta_2|b\rangle)
+\frac{1}{\sqrt{2}}(\alpha(r+t_0)|R^{r,D_2}\rangle+\beta(r+t_0)|R^{l,D_1}\rangle \nonumber\\&&+\alpha\sqrt{1-(t-t_0)^2}|L^{r,D_4}\rangle-\beta\sqrt{1-(t-t_0)^2}|L^{l,D_3}\rangle)
 \nonumber\\&&\otimes(\gamma|\uparrow\rangle+\eta|\downarrow\rangle)\otimes(\delta_1|a\rangle+\delta_2|b\rangle).
\end{eqnarray}
If $D_1$, $D_2$, $D_3$ and $D_4$ are not clicked, the system will collapse to
\begin{eqnarray}                  \label{eq31}
|\varphi\rangle_6 &=&\gamma(t-t_0)(\alpha|R^l\rangle+\beta |L^l\rangle)(\delta_1|a^l\rangle+\delta_2|b^l\rangle)|\uparrow\rangle \nonumber\\&&
                             -\eta(t-t_0)(\alpha|R^r\rangle +\beta|L^r\rangle)(\delta_1|a^r\rangle+\delta_2|b^r\rangle)|\downarrow\rangle.
\end{eqnarray}

From Eqs. (\ref{eq25})-(\ref{eq31}), one can see that Fig. \ref{Router-polarization} accomplished a hyper-router acting on the polarization and spatial DOFs. The signal photon can be directed to right port or the left port controlled by the spin of the electron in QD.


\section{Hyperparallel DRAM via practice QD-cavity emitter}\label{sec4}

\begin{figure}[htp]
\centering
\includegraphics[width=8.5 cm,angle=0]{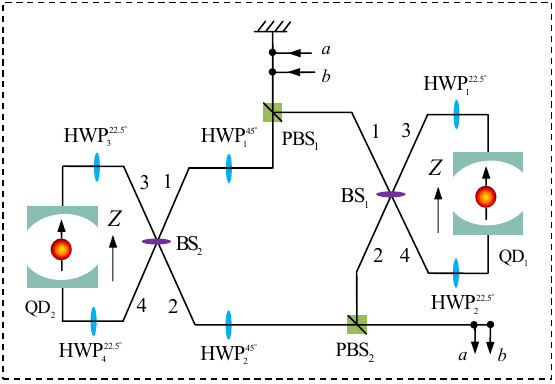}
\caption{ Schematic diagram of hyper-DRAM.}
\label{DRAM-polarization}
\end{figure}

DRAM is the key quantum technology for quantum computers and quantum networks, which can load, store, and unload  polarization photon controlled by the state of the control qubits. Fig. \ref{DRAM-polarization} shows the diagram of an optical spin-based hyper-DRAM, and the loading, storing, and reading out of the photon are controlled by BS$_1$ and BS$_2$. Considering the initialization state of the system constituted of two QDs and one photon is prepared as
\begin{eqnarray}                  \label{eq32}
|\phi\rangle_0&=&(\alpha|R\rangle+\beta|L\rangle)\otimes(\delta_1|a\rangle+\delta_2|b\rangle)
\otimes(\gamma_1|\uparrow_1\rangle+\gamma_2|\downarrow_1\rangle)\otimes(\eta_1|\uparrow_2\rangle+\eta_2|\downarrow_2\rangle).
\end{eqnarray}
Here, $|\alpha|^2+|\beta|^2=1$, $|\delta_1|^2+|\delta_2|^2=1$, $|\gamma_1|^2+|\gamma_2|^2=1$, and $|\eta_1|^2+|\eta_2|^2=1$.

First, PBS$_1$ splits the input photon state $(\alpha|R\rangle+\beta|L\rangle)\otimes(\delta_1|a\rangle+\delta_2|b\rangle)$ into two components, $\alpha|R\rangle(\delta_1|a\rangle+\delta_2|b\rangle)$ and $\beta|L\rangle(\delta_1|a\rangle+\delta_2|b\rangle)$. And then, a bit-flip operation $\sigma_x$ is performed on the $R$-polarized component by using HWP$^{45^\circ}_1$. That is, PBS$_1$ and HWP$^{45^\circ}_1$ transform the initial state $|\phi\rangle_0$ into
\begin{eqnarray}                  \label{eq33}
|\phi\rangle_1=(\alpha|L^{l,1}\rangle+\beta|L^{r,1}\rangle)\otimes(\delta_1|a\rangle+\delta_2|b\rangle)
\otimes(\gamma_1|\uparrow_1\rangle+\gamma_2|\downarrow_1\rangle)\otimes(\eta_1|\uparrow_2\rangle+\eta_2|\downarrow_2\rangle).
\end{eqnarray}

Second,  after the $L^{r,1}$ ($L^{l,1}$) component interacts with the block constituted of BS$_1$,  HWP$^{22.5^\circ}_1$, HWP$^{22.5^\circ}_2$, and QD$_1$ (BS$_2$,  HWP$^{22.5^\circ}_3$, HWP$^{22.5^\circ}_4$, and QD$_2$), the  state of the system can be written as
\begin{eqnarray}                  \label{eq34}
|\phi\rangle_2=-(\alpha|L^{l,1}\rangle+\beta|L^{r,1}\rangle)\otimes(\delta_1|a\rangle+\delta_2|b\rangle)
\otimes(\gamma_1|\uparrow_1\rangle+\gamma_2|\downarrow_1\rangle)\otimes(\eta_1|\uparrow_2\rangle+\eta_2|\downarrow_2\rangle).
\end{eqnarray}
Here BS$_1$ and BS$_2$ complete the transformations
\begin{eqnarray}                  \label{eq35}
&&|L^1\rangle\leftrightarrow\frac{1}{\sqrt{2}}(|L^{3}\rangle+|L^{4}\rangle), \quad
|L^2\rangle\leftrightarrow\frac{1}{\sqrt{2}}(|L^{3}\rangle-|L^{4}\rangle), \nonumber\\&&
|L^{3}\rangle\leftrightarrow\frac{1}{\sqrt{2}}(|L^{1}\rangle+|L^{2}\rangle), \quad
|L^{4}\rangle\leftrightarrow\frac{1}{\sqrt{2}}(|L^{1}\rangle-|L^{2}\rangle).
\end{eqnarray}

Third, HWP$^{45^\circ}_1$ induces $L^{l,1}$ to be $R^{l,1}$. And then $R^{l,1}$ and $L^{r,1}$ pass though PBS$_1$ and arrive at a high reflective mirror.  Subsequently, the mirror reflects the photon into the second round, after  $N$ rounds, the state of the whole system is evolved as
\begin{eqnarray}                  \label{eq36}
|\phi\rangle_3&=&(-1)^N(\alpha|L^{l,1}\rangle+\beta|L^{r,1}\rangle)\otimes(\delta_1|a\rangle+\delta_2|b\rangle)
\otimes(\gamma_1|\uparrow_1\rangle+\gamma_2|\downarrow_1\rangle) \nonumber\\&&\otimes(\eta_1|\uparrow_2\rangle+\eta_2|\downarrow_2\rangle).
\end{eqnarray}

From Eqs. (\ref{eq32})-(\ref{eq36}), one can see that the photons are loaded and stored.

For reading out, after the wave packets interact with the two QDs, we rotate BS$_1$ and BS$_2$ by 180$^\circ$ to complete the transformations
\begin{eqnarray}                  \label{eq37}
&&|L^1\rangle\rightarrow\frac{1}{\sqrt{2}}(-|L^3\rangle + |L^4\rangle),\quad
|L^2\rangle\rightarrow\frac{1}{\sqrt{2}}(|L^3\rangle + |L^4\rangle), \nonumber\\&&
|L^3\rangle\rightarrow\frac{1}{\sqrt{2}}(-|L^1\rangle + |L^2\rangle),\quad
|L^4\rangle\rightarrow\frac{1}{\sqrt{2}}(|L^1\rangle + |L^2\rangle).
\end{eqnarray}
After the wave packets mix at BS$_1$ and BS$_2$, the state of the system is changed to be
\begin{eqnarray}                  \label{eq38}
|\phi\rangle_4&=&(-1)^{N+1}(\alpha|L^{l,2}\rangle+\beta|L^{r,2}\rangle)\otimes(\delta_1|a\rangle+\delta_2|b\rangle)
\otimes(\gamma_1|\uparrow_1\rangle+\gamma_2|\downarrow_1\rangle) \nonumber\\&&\otimes(\eta_1|\uparrow_2\rangle+\eta_2|\downarrow_2\rangle).
\end{eqnarray}
Next, as shown in Fig. \ref{DRAM-polarization}, $L^{l,2}$-polarized component will be transformed into $R^{l,2}$-polarized component by HWP$^{45^\circ}_2$. That is, HWP$^{45^\circ}_2$ evolves $|\phi\rangle_4$ to
\begin{eqnarray}                  \label{eq39}
|\phi\rangle_5&=&(-1)^{N+1}(\alpha|R^{l,2}\rangle+\beta|L^{r,2}\rangle)\otimes(\delta_1|a\rangle+\delta_2|b\rangle)
\otimes(\gamma_1|\uparrow_1\rangle+\gamma_2|\downarrow_1\rangle) \nonumber\\&&\otimes(\eta_1|\uparrow_2\rangle+\eta_2|\downarrow_2\rangle).
\end{eqnarray}
Finally, $R^{l,2}$-polarized and $L^{r,2}$-polarized components go out by PBS$_2$. That is, the information of the photon is read out.

\section{Conclusion}

The realizations of quantum computers, quantum networks, and quantum internet, require not only  quantum gates and quantum memories, but also demand quantum transistors, routers, and DRAMs. Optical transistor can be used for a bridge between quantum networks and all-optical networks. The source light beam (strong light) in the all-optical transistor is controlled by the gate photon (weak light). It is known that the more complex quantum networks are, the more pronounced is, the need for  directing the signal qubit (inputs) to its intended destination (outputs) according to the state of the control qubit. Optical DRAM can be used for loading, storing, and reading out of photons.

The previous works about optical transistor and other optical devices mainly are limited to one DOF case \cite{OFC2,transistor,wang-router,OFC6,OFC7,SPT5}. In this paper, we designed compact quantum circuits for determinately implementing single photon hyper-transistor, hyper-router and hyper-DRAM, respectively. The strong interactions between individual photons could be achieved by employing cavity quantum electrodynamics system with QD. Our schemes act on the polarization and the spatial DOFs of the photon, simultaneously, and single photon hyper-transistor can be applied to amplify an arbitrary single-photon hyperparallel state to the same $N$-photon hyperentanglement state.

The balanced reflectance for the ``hot'' and the ``cold''  cavity, which is necessary for the single-sided cavity, but is not necessary for our schemes to get high fidelities.  In previous works \cite{repeater,hyper-purification1,hyper-purification2,hypergate2,OVE1,DM,QC2,QC3}, the imperfect birefringence of the cavity induced by side leakage of optical cavity are often not taken into account. These imperfections reduce the fidelity of the practical emitter by a few percents.
Our schemes  not only are robust against the imperfect birefringence of the cavity, and the unity fidelities can be achieved but also are immune to the quantum fluctuations \cite{fluctuations} (because the QD stays in the ground state), spin noises, spectral diffusion, and pure dephasing in QDs. Moreover, the strong coupling limitation (the fidelity climbs with QD-cavity coupling strength increasing \cite{Hu2}) is avoided and based on the hyperparallel programs, high capacity, high speed, low loss rate characters can be come true in our work.

It is known that strong coupling is a challenge in experiment. Fortunately, $\kappa_s/\kappa=0.7$ with $g/(\kappa+\kappa_s)=1.0$ was reported in micropillar cavity in 2007 \cite{strong}. In 2008, coupling strength  was raised from $g/(\kappa +\kappa_s)=0.5$ ($Q$ = 8800) to $g/(\kappa +\kappa_s)=2.4$ \cite{arised}. $\kappa_s/\kappa=0.05$ in the strong regime has been achieved in a pillar microcavity with $Q =9000$ \cite{achieved}. Polarization degenerate cavity with a single QD for polarization-based QIP has been achieved in experiment \cite{degenerate1,degenerate2}.
The mechanism of our schemes is deterministic and unity in principle. The efficiencies of our schemes are not unity. The success of our schemes are heralded by single photon detectors. Moreover, the success probabilities of our schemes are also effected by $\kappa_s$, $g/\kappa$,  PBS, VBS, and BS. Some inevitable experimental imperfections, including the effects of the hole mixing, the dark transitions, the single photon detector dark counts, and the balancing of PBSs and BSs,  will decrease the fidelities and the efficiencies of the presented optical elements.


\section*{Acknowledgments}

The work is supported by the National Natural Science Foundation of China under Grant No. 11604012, and the Fundamental Research Funds for the Central Universities under Grant No. 230201506500024, the National Natural Science Foundation of China under Grant No. 11647042, and the Fundamental Research Funds for the Central Universities under Grant No. FRF-BR-17-004B.

%
\bibliography{sample}
%
%
%

\end{document}